%% file: main.tex
\documentclass[sigconf]{acmart}

\usepackage{xcolor}
\usepackage{amsmath}
\usepackage{multirow}
\usepackage{array}
\usepackage{tabularx}
\usepackage{hyperref}
\usepackage[linesnumbered,ruled,vlined]{algorithm2e}
\usepackage[noend]{algpseudocode}
\usepackage{placeins}
\usepackage{colortbl}
\usepackage{siunitx}
\usepackage{makecell}

\AtBeginDocument{%
  \providecommand\BibTeX{{%
    \normalfont B\kern-0.5em{\scshape i\kern-0.25em b}\kern-0.8em\TeX}}}

\setcopyright{acmlicensed}
\copyrightyear{2024}
\acmYear{2024}
\acmDOI{XXXXXXX.XXXXXXX}

\acmConference[arxiv]{}{}

\acmISBN{978-1-4503-XXXX-X/18/06}

\begin{document}

\title{RiskMiner: Discovering Formulaic Alphas via Risk Seeking Monte Carlo Tree Search}

\author{Tao Ren}
\affiliation{%
    \institution{Wuhan Institute of Artificial Intelligence, Guanghua School of Management, Peking University}
    \institution{Xiangjiang Laboratory}
    \country{China} 
}
\email{rtkenny@stu.pku.edu.cn}

\author{Ruihan Zhou}
\affiliation{%
    \institution{Wuhan Institute of Artificial Intelligence, Guanghua School of Management, Peking University}
    \institution{Xiangjiang Laboratory}
    \country{China} 
}
\email{rhzhou@stu.pku.edu.cn}

\author{Jinyang Jiang}
\affiliation{%
    \institution{Wuhan Institute of Artificial Intelligence, Guanghua School of Management, Peking University}
    \institution{Xiangjiang Laboratory}
    \country{China} 
}
\email{jinyang.jiang@stu.pku.edu.cn}

\author{Jiafeng Liang}
\affiliation{%
  \institution{Harbin Institute of Technology}
  \city{Harbin}
  \country{China}
}
\email{jfliang@ir.hit.edu.cn}

\author{Qinghao Wang}
\affiliation{%
  \institution{Peking University}
  \city{Beijing}
  \country{China}}
\email{qinghw@pku.edu.cn}

\author{Yijie Peng}
\authornote{Corresponding authors.}
\affiliation{%
    \institution{Wuhan Institute of Artificial Intelligence, Guanghua School of Management, Peking University}
    \institution{Xiangjiang Laboratory}
    \country{China} 
}
\email{pengyijie@pku.edu.cn}

\begin{abstract}
The formulaic alphas are mathematical formulas that transform raw stock data into indicated signals. In the industry, a collection of formulaic alphas is combined to enhance modeling accuracy. Existing alpha mining only employs the neural network agent, unable to utilize the structural information of the solution space. Moreover, they didn't consider the correlation between alphas in the collection, which limits the synergistic performance. To address these problems, we propose a novel alpha mining framework, which formulates the alpha mining problems as a reward-dense Markov Decision Process (MDP) and solves the MDP by the risk-seeking Monte Carlo Tree Search (MCTS). The MCTS-based agent fully exploits the structural information of discrete solution space and the risk-seeking policy explicitly optimizes the best-case performance rather than average outcomes. Comprehensive experiments are conducted to demonstrate the efficiency of our framework. Our method outperforms all state-of-the-art benchmarks on two real-world stock sets under various metrics. Backtest experiments show that our alphas achieve the most profitable results under a realistic trading setting.
\end{abstract}



\maketitle

\section{Introduction}
Alpha factors or alphas transform raw stock data into indicated signals for future stock trends. There are two types of alphas: formulaic alphas\cite{autoalpha,huataigp1,huataigp2,alphaevolve,alphagen} and machine-learning alphas\cite{adaptive_TF,hierarchical_TF,doubleadapt,dva,sfm,hist}. The formulaic alphas\cite{findingalpha}, expressed by mathematical formulas, perform feature engineering on the price/volume data. The formulaic alphas can be used either as trading signals or as input for complex machine-learning forecasting models. The machine learning alphas refer to machine learning models designed to produce trading signals\cite{factorvae}. Even though they usually are more predictive than formulaic alphas, the black-box nature renders the lack of interpretability.




\begin{figure*}[!t]
  \includegraphics[width=\textwidth]{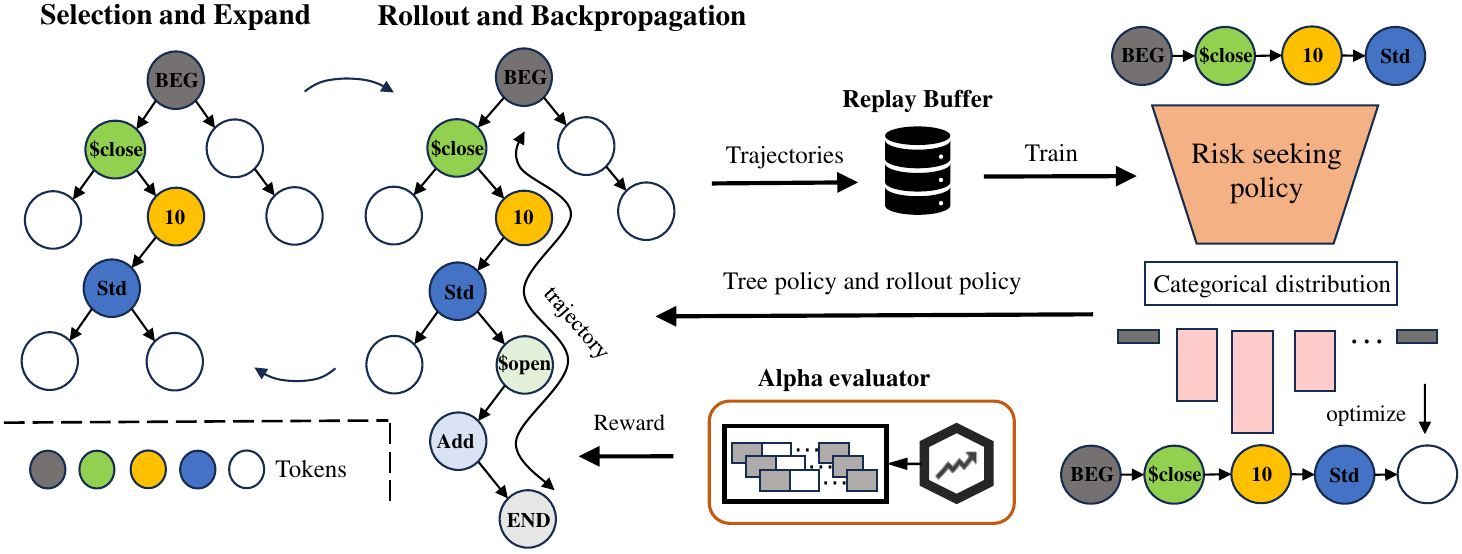}
  \caption{Algorithm framework diagram. The alpha evaluator provides reward signals for the reward-dense MDP. On the left side, the MCTS serves as a sampler to sample trajectories from the MDP. On the right side, a risk-seeking policy is trained and the policy will be used as the tree policy and rollout policy in the MCTS.}
  \label{fig:algdiagram}
\end{figure*}

Recently, researchers have designed various frameworks \cite{alphagen,alphaevolve,autoalpha,huataigp2,huataigp1} to generate the formulaic alpha automatically. Most of the alpha generation methods are based on genetic programming (GP) \cite{alphaevolve,autoalpha,huataigp2,huataigp1}. By performing bionic operations on the expression tree, new alphas are generated based on existing alphas. Alphagen\cite{alphagen} uses Reinforcement Learning (RL) to generate alphas synergistically. They formulate the mining pipeline as a Markov decision process (MDP) and use proximal policy optimization (PPO) \cite{PPO} to solve it. Even though they have achieved better performance than GP, their mining pipeline still has many limitations. First, the reward-sparse nature of the MDP makes the learning process highly non-stationary, as it provides limited feedback for the search process. Second, similar to chess games, the discrete solution space for the formulas is huge and PPO is not the optimal method to explore such a solution space. Third, the alphagen only considers the average performance when mining alpha whereas the best-case performance should be the focus when mining alpha factors. To overcome these challenges, we introduce RiskMiner, a novel alpha mining framework addressing all the limitations.

Figure \ref{fig:algdiagram} illustrates the main idea of our framework. The MCTS serves as a sampler providing training data for risk-seeking policy optimization. Meanwhile, the trained risk-seeking policy network will be used in the MCTS search cycle. The two modules operate alternately as an alpha mining agent. 

To solve the first limitation, our approach involves a reward-dense MDP, which is structured with both intermediate and terminal rewards, thereby providing more frequent and informative feedback throughout the learning process. The mining agent can discover "excellent and different" synergistic alphas from solving the reward-dense MDP.

To solve the second limitation, we employ Monte Carlo Tree Search (MCTS), which is a potent tool for solving discrete sequential decision-making problems, records the explored solution space through its tree-like structure, and searches unexplored areas based on specific policies. MCTS plays a pivotal role in the champion-defeating algorithm, e.g. AlphaZero\cite{alphazero} and MuZero\cite{muzero}. The success in playing Go and other board games has demonstrated the potential of MCTS for solving discrete planning problems in other fields. Since then, there has been a series of research applying the MCTS framework to solve other real-world problems including discovering faster sorting algorithms\cite{alphadev} and matrix multiplication algorithms \cite{alphatensor}. While the MCTS has been successfully applied in various domains, to our best knowledge, \emph{this is its first application in discovering alphas in quantitative finance.}

Unlike other reinforcement learning tasks where mean or worst-case performance is typically prioritized (such as game playing\cite{alphago_zero}, robotic control\cite{yyd}, and portfolio management\cite{finrl}), alpha mining should focus on best-case performance. The intuitive explanation for this is that \emph{bumping into bad formulas won't hurt whereas discovering a good formula will significantly boost the overall performance.} After sampling trajectories from the MCTS, we employ a risk-seeking policy optimization method to optimize the best-case performance, addressing the third limitation.

We evaluate our framework on real-world stock data from two stock sets: constituent stocks from CSI300 and CSI500. We compare our framework with a couple of baselines under various evaluation metrics. Our method outperforms all the benchmarks in every metric. We also conduct an investment simulation study under a realistic trading setting. Employing a simple and widely used trading strategy, our methods achieve the most profitable outcomes among all compared methods. In the end, we design an ablation study to investigate the contribution of each component. 

The contributions of our work can be summarized as follows:
\begin{itemize}
    \item We design a new reward-dense MDP to stabilize the synergistic alpha mining process.
    \item We propose RiskMiner, a novel alpha mining framework that combines MCTS and risk-seeking policy, to efficiently search the solution space.
    \item We conduct the experiments on signal-based metrics and investment simulation on the real-world data. An ablation study is also included. The results verified our RiskMiner framework’s superiority and validity.
\end{itemize}

\section{Related Work}

\textbf{Formulaic alphas.} The expression space for formulaic alphas is extremely large, owing to the myriad of operands and operators involved in their construction. GP has been the mainstream method for discovering formulas. \cite{huataigp1} and \cite{huataigp2} modify the gplearn library by incorporating non-linear and time-series operators, allowing for a more nuanced exploration of the alpha expression space. Autoalpha\cite{autoalpha} enhances GP by integrating Principal Component Analysis (PCA), which aids in steering the search away from already explored solution spaces. Alphaevolve \cite{alphaevolve} includes vector and matrix operator in their evolutionary framework to further enhance the predictive ability of the alphas. However, complicated vectors and matrix operations lead to reduced interpretability. Alphagen \cite{alphagen} formulates the mining pipeline as an MDP and uses PPO to solve it. Even though Alphagen has surpassed previous evolving-based methods, its deficiency is obvious. Due to the sparse reward characteristic, the MDP is highly non-stationary thereby making the learning process difficult. Meanwhile, the PPO agent, with only the neural-network-based agent, doesn't efficiently exploit the structural information of the search space.

\textbf{Stock trend forecasting.} 
Using end-to-end machine-learning models to forecast stock trends has attracted enormous attention from researchers. The vanilla idea is to use time-series models\cite{transformer} to predict future trends. \cite{hierarchical_TF} employs the transformer to forecast future stock trends. \cite{adaptive_TF} applies orthogonal regularization tricks to make multi-head attention more efficient in extracting hierarchical information from stock data. Inspired by signal processing theory, \cite{sfm} proposes a DFT-like forecasting method to capture the latent trading pattern underlying the fluctuation of stock price. Researchers also attempt to use diffusion models to model the distribution of stock returns. \cite{dva} combines the diffusion model and variational autoencoder to regress future returns. The machine-learning methods usually have strong forecasting abilities but they require high-quality features constructed by formulaic alphas. 

Since real-world events are also correlated with stock price movement, forecasting methods that utilize textual inputs, like news articles\cite{knowledge-stock, newslistening_stock}, are gaining attention. \cite{RL_NQT} develops a news-driven trading algorithm, which can detect abrupt jumps caused by spotlight news. The burgeoning field of Large Language Models (LLMs) presents new research frontiers, with emerging research\cite{fingpt,finmem} exploring their potential to provide financial insights. 

\section{PROBLEM FORMULATION}
\subsection{Formulaic Alpha}
Formulaic alpha is defined as a mathematical expression $f(\cdot)$. The formula transforms the original price data $X_t$ of stocks and other relevant financial and market data into alpha values $z_t=f(X_t) \in \mathbb{R}^n$, in which $n$ is the number of stocks. We use the term alpha to refer to the formula and the calculation results. Good alpha should have a significant correlation between the current alpha values and the subsequent stock returns.
The Information Coefficient (IC) is a popular metric in the industry for evaluating alphas. The IC is the Pearson correlation between the alpha value $z_t$ and the future stock returns $r_{t+1}$, expressed by the formula 
\begin{equation}
    \mathrm{IC}=\frac{\mathrm{Cov}(z_t,r_{t+1})}{\sigma_{z_t}\sigma_{r_{t+1}}},
\end{equation}
in which $\mathrm{Cov}$ is the covariance and $\sigma$ is the standard deviation.

When the alpha value and the returns are ranked by their magnitude, the resulting IC is referred to as RankIC. The purpose of this approach is to mitigate the impact of extreme values and to emphasize the relative order of the predicted stocks. For instance, an original series of alpha values such as (0.1, 0.2, 0.3) becomes (1, 2, 3) after ranking, which is then used to calculate the RankIC:
\begin{equation}
\mathrm{RankIC}=\frac{\mathrm{Cov}(\mathrm{rank}(z_t),\mathrm{rank}(r_{t+1}))}{\sigma_{\mathrm{rank}(z_t)}\sigma_{\mathrm{rank}(r_{t+1})}}.
\end{equation}

The range of IC values extends from -1 to 1, with values closer to 1 or -1 indicating a stronger predictive capability of the alpha.
The goal of alpha mining is to discover alpha expressions with strong predictive ability, reflected in higher IC and RankIC values. Additionally, the similarity between two alphas can be assessed by calculating the Pearson correlation coefficient between them, known as the Mutual Information Coefficient (mutIC), which is used to evaluate the similarity between two alphas in predicting the same target:
\begin{equation}
    \mathrm{mutIC}=\frac{\mathrm{Cov}(z_t,z_t^*)}{\sigma_{z_t}\sigma_{z_t^*}}.
\end{equation}
\subsection{Mining Combinational Alpha}
It is a common practice to use an alpha synthesis model to generate a composite alpha value from a group of alphas. The composite alpha can better guide the construction of the investment portfolio. Let $\mathcal{F}=\{f_1, f_2,\dots,f_k\}$ be a set containing $k$ alphas.  the alpha synthesis model, $c(\cdot|\mathcal{F},\omega)$, integrates the information of each alpha. Given the model parameters, $\omega$, the composite alpha value $z_t=c(X_t|\mathcal{F},\omega)$ can be computed.

The composite alpha, integrating diverse information from multiple alphas, provides a comprehensive perspective of the market and often has strong predictive capabilities. In the alpha mining procedure, the objective is to identify not only alphas with high IC but also those that can further enhance the composite alpha's IC when combined with other alphas, achieving synergistic effects.

Discovering an ideal set of alphas is a highly challenging task. An ideal alpha combination should meet the following criteria: \emph{the alphas in the set should have high individual IC, meanwhile maintaining mutIC as low as possible with other alphas.} High mutIC between alphas indicates market information overlaps. The redundancy within the alpha pool leads to limited improvement in the composite alpha's performance. Therefore, finding a combination of "excellent and unique" alphas is essential in alpha mining.

\subsection{Reverse Polish Notation}
\begin{figure}[h]
  \centering
  \includegraphics[width=0.6\linewidth]{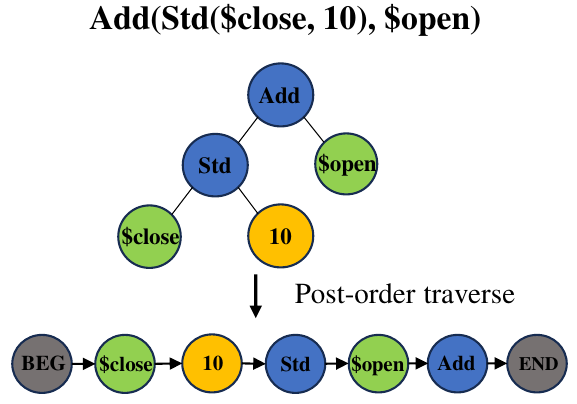}
  \caption{Reverse Polish Notation(RPN).}
  \label{fig:RPN}
\end{figure}

In our approach, we represent alphas using Reverse Polish Notation (RPN), setting up the stage for the subsequent design of an MDP. Reverse Polish Notation is derived from the post-order traversal of an expression binary tree. With operands as leaf nodes and operators as non-leaf nodes, the expression binary tree unambiguously represents the expression of an alpha. The RPN models the formula as a sequence of tokens. Figure \ref{fig:RPN} provides an example of the RPN.

\section{METHODOLOGY}
To mine the "excellent and different" alpha set, we first formulate a reward-dense MDP that enables the search algorithm to efficiently learn the characteristics of the expression space. Then we propose a novel risk-seeking MCTS to explore the expression space. To be specific, MCTS and risk-seeking policy optimization are executed alternately to mining alphas.

\subsection{Alpha Pool}
We typically utilize an alpha pool to synthesize a group of alphas. An ideal alpha pool should be a structurally simple and efficient machine learning model, which can be either a linear model or a tree model. Given the efficacy, interpretability, and simplicity, we opt for using a linear model for alpha synthesis.

Given the alpha pool model $c(\cdot|\mathcal{F},\omega)$, we calculate the composite alpha $z_t=c(X_t|\mathcal{F},\omega)$ through a weighted synthesis approach. Assuming there are $k$ alphas in the pool, the model parameters $\omega=(\omega_1,\omega_2,\dots,\omega_k)$ represent the weight coefficients for each alpha. The absolute values of the elements in the weight vector reflect the importance of the corresponding alpha. $f(X_t)=(f_1(X_t),f_2(X_t),\dots,f_k(X_t))$ is the current value of the $k$ alphas, and $z_t=\omega\cdot f(X_t)$ represents the value of the composite alpha.

During the model training process, we employ Mean Squared Error (MSE) as the loss function to measure the gap between the synthesized alpha and the future stock returns:
\begin{equation}
    \mathcal{L}(\omega)=\frac{1}{nT}\sum_{t=1}^T||z_t-r_{t+1}||^2.
\end{equation}
By using gradient descent to minimize the loss, we obtain the optimal weight $\omega$ for the current alpha set $\mathcal{F}$. The pool size $K$ is predefined. The new alpha will be added to the pool incrementally. When a new alpha is added to the pool, gradient descent is performed to get the updated weight of the $k+1$ alphas. If the number of alphas has reached the threshold $K$, the least principal alpha, whose absolute weight value is the smallest, is removed from the pool. The pseudocode for maintaining the alpha pool is shown in Algorithm \ref{alg:alphapool}.

\subsection{Reward-Dense MDP}
We construct a reward-dense MDP for the alpha mining problem. By solving this MDP, the algorithm can identify alphas exhibiting high IC performance.

\emph{Key Concept in the MDP — Token:} A Token is the fundamental building block in this MDP, representing an abstraction of operands and operators.  Within this framework, the state of the MDP, $s_t$, is defined as the current sequence of selected Tokens, while the action, $a_t$, is to select the next Token. The transition is deterministic. Each decision sequence (episode) starts with a specific beginning Token (BEG) and concludes upon selecting the ending Token (END).

\emph{Design of Intermediate State Rewards:} Intermediate rewards are set for states that have not reached the end state (i.e., have not selected the END Token). If the current Token sequence forms a valid RPN expression, we can compute the IC value of the alpha formed by this Token sequence.  Furthermore, if the alpha pool is non-empty, we calculate the mutIC between this Token sequence and each alpha in the alpha pool. The intermediate reward signal is derived from these calculations:
\begin{equation}
    \mathrm{Reward_{inter} }=\mathrm{IC}-\lambda\frac{1}{k}\sum_{i=1}^k \mathrm{mutIC}_i,
\end{equation}
where $k$ is the number of existing alphas in the pool, $\mathrm{mutIC}_i$ denotes the mutual IC value between the current alpha and the $i$-th alpha in the pool, and $\lambda$ is a hyperparameter in the MDP, set to 0.1 in this study.
\begin{figure}[h]
  \centering
  \includegraphics[width=\linewidth]{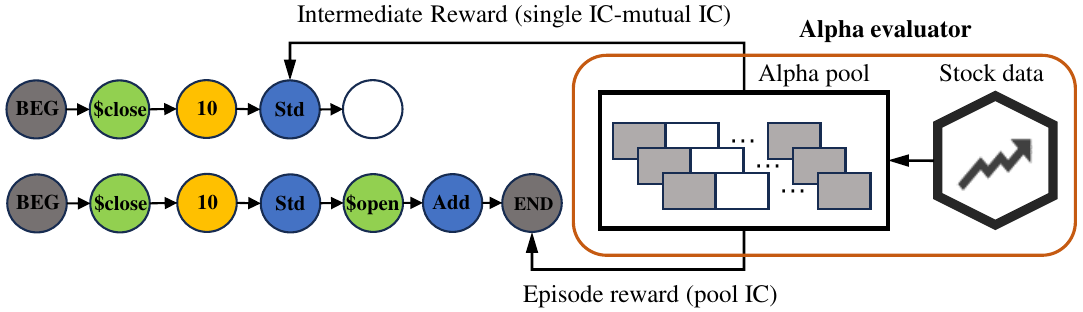}
  \caption{Reward-Dense MDP. The intermediate reward is designed for the legal but not complete expression. When the episode is terminated, an episode reward is assigned.}
\end{figure}

\emph{Episode Termination and Reward Allocation:} The selection of the END Token signifies the end of an episode.  At this point, the corresponding alpha is added to the alpha pool, and a specific algorithm (Algorithm 1) is executed to obtain the current composite alpha value.  Ultimately, the IC value of the composite alpha serves as the overall reward $\mathrm{Reward_{end}}$ for that episode. The maximum length of the episode is 30.

\subsection{Risk-based Monte Carlo Tree Search}
\begin{figure*}[!t]
  \includegraphics[width=0.9\textwidth]{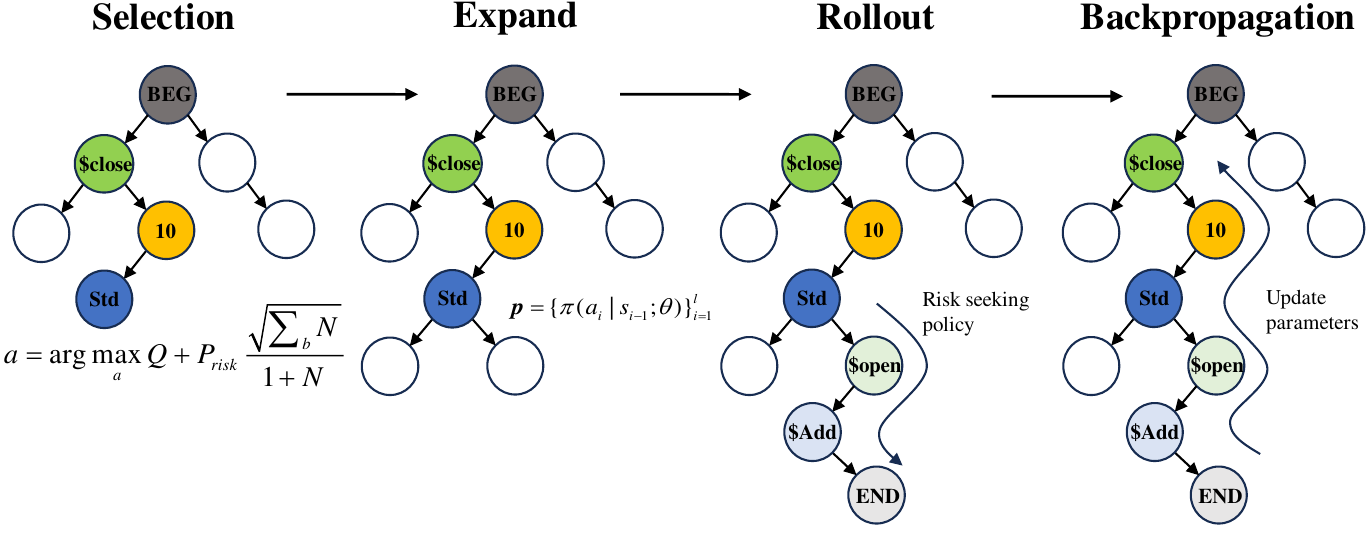}
  \caption{Search Cycle of Monte Carlo Tree Search. Selection: start from the root and reach the leaf according to the Tree Policy. Expand: expand the leaf node and assign the initial value for the newly added edges. Rollout: reach the terminated state with the Rollout Policy. Backpropagation: update the value of the edge along the trajectory with the intermediate and terminated reward.}
  \label{fig:mcts}
\end{figure*}

To effectively search the expression space, we design a special MCTS, aimed at efficiently exploring the vast and discrete alpha solution space composed of RPN expressions. Figure \ref{fig:mcts} shows a single cycle of MCTS. A single search cycle in MCTS consists of four phases: Selection, Expansion, Rollout, and Backpropagation. In the Selection phase, the Tree Policy is a crucial mechanism guiding the search process from the root to the leaf node, determining which leaf node is worthy of further exploration. The Rollout phase aims to assess the value of the current leaf node by the Rollout Policy. The efficacy of the Tree Policy and Rollout Policy directly impact the quality of the search.


To enhance the efficiency of searching the expression space, this study trains a \textbf{policy network} via a novel risk-seeking policy gradient method. \emph{The policy generated by the network is used as both the Tree Policy and Rollout Policy to assist in the selection and rollout processes.} The training method of this network will be detailed in Section 4.4.

Each edge $(s,a)$ in the Monte Carlo tree contains the following information: $\{N(s,a),P(s,a),Q(s,a),R(s,a)\}$, where $N(s,a)$ is the number of visits to the edge, $P(s,a)$ is the prior probability of the edge as given by the risk policy network, and $Q(s,a)$ is the value of the edge.

\emph{Selection and Expansion:} In the Monte Carlo Tree, nodes represent the tokens of the expression, and edges represent the action of selecting the next token from the current token sequence. The root node of the tree is the BEG Token, where the search begins. During the selection process, we employ the following PUCT formula to determine the next non-leaf node to select:
\begin{equation} \label{eq:PUCT}
    a_t=\arg\max_a Q(s,a)+P(s,a)\frac{\sqrt{\sum_bN(s,b)}}{1+N(s,a)}.
\end{equation}
Upon reaching a leaf node that is not an endpoint (i.e., not the END Token), we invoke the risk policy network, $p_l=\pi_{risk}(s_l)$, and assign initial values to the newly expanded edges$\{N(s,a)=0,P(s,a)=p_l,Q(s,a)=0,R(s,a)=0\}$. Each time an intermediate node is selected, if the current token sequence is valid, then the reward of edge $(s,a)$ is updated with $R(s,a)=r_t=\mathrm{Reward_{inter}}$. 

\emph{Rollout and Backpropagation:} After completing the selection and reaching a non-terminal leaf node, the rollout is performed. During the rollout, the next action is sampled from the probability given by the policy network until the END Token is met. The intermediate rewards $\mathrm{Reward_{inter}}$ generated during the rollout, along with the final reward $\mathrm{Reward_{end}}$, are summed to get the node value estimation $v_l$. For non-leaf node $k=l,\dots,0$, we perform a $l-k$ step bootstrap and get the following cumulative rewards:
\begin{equation}
    G_k=\sum_{i=0}^{l-1-k}\gamma^ir_{k+1+i}+v_l,
\end{equation}
where $\gamma$ is set to 1 in this scenario because we encourage the exploration of long expressions.
For $k=l,\dots,1$, we update the data on the edge $(s_{k-1},a_k)$:
\begin{equation}
    Q(s_{k-1},a_k)=\frac{N(s_{k-1},a_k)\times Q(s_{k-1},a_k)+G_k}{N(s_{k-1},a_k)+1},
\end{equation}
\begin{equation}
    N(s_{k-1},a_k)=N(s_{k-1},a_k)+1.
\end{equation}
In one search loop of MCTS, the process from the BEG Token to the END Token constitutes a complete episode trajectory
$$\tau=\{s_0,a_1,r_1,s_1,\dots,s_{T-1},a_T,r_T, s_T\}.$$ Episodes generated during the search process are stored in the replay buffer. When the number of episodes in the buffer reaches a predetermined threshold, the search is stopped, and these data are used to train the risk policy network for subsequent search cycles.
\subsection{Risk-Seeking Policy Optimization}
Mainstream RL algorithms typically aim at maximizing the expectation of cumulative rewards, which can be ineffective in optimizing the best-case performance of the policy. We design a novel policy-based RL algorithm to train a risk-seeking policy network that prioritizes identifying optimal alphas by focusing on best-case scenarios. In standard planning and decision-making problems, tail risks (i.e., the worst-case performance of a policy) are often emphasized, to minimize losses in the worst-case scenarios. However, in the context of alpha search, our focus shifts towards pursuing the best-case performance. We utilize quantile optimization, an effective mathematical tool for reshaping the distribution, to tailor our risk-seeking policy. \emph{The trained risk-seeking policy serves as the rollout and tree policy in the MCTS to optimize the best-case performance during the alpha search process.} Through this approach, we can conduct more efficient searches across the expression space.

In each episode, a trajectory $\tau$ is generated by following the action selecting policy $\pi(\cdot|\cdot;\theta)$, which is represented by a neural network with parameter $\theta$. The corresponding cumulative reward is given by $R(\tau)=\sum_{t=0}^{T-1}\gamma^t r_t$, which follows a CDF $F_R(\cdot;\theta)$. Classical policy-based RL aims to optimize the expectation objective $J(\theta)=\mathbb{E}_{\tau\sim \Pi(\cdot;\theta)}[R(\tau)]$. However, our interest lies in the quantile of the cumulative reward, which can be defined as
$q(\theta;\alpha)=F_R^{-1}(\alpha;\theta)$ when $F_R(\cdot;\theta)$ is continuous. We are concerned with the upper $\alpha$-quantile of $R(\tau)$, thus solving the optimization problem
\begin{equation}
    \max_{\theta\in \Theta}J_{risk} (\theta;\alpha) = \max_{\theta\in \Theta} q(\theta;1-\alpha),
\end{equation}
where $\Theta$ denotes the parameter space.
The policy optimization procedure under the quantile criterion can be implemented by two coupled recursions. We first need to estimate the quantile of the cumulative reward. 
The upper $\alpha$-quantile is tracked by 
\begin{equation} \label{eq:quantile_update}
    q_{i+1}=q_i+\beta(1-\alpha-\mathbf{1}\{R(\tau_i)\leq q_i\}),
\end{equation}
which is a numerical approach for solving the root searching problem $F_R(q(1-\alpha;\theta);\theta) = 1-\alpha$.
Then we can perform gradient ascent with the direction calculated by the following theorem. 
\begin{theorem} \label{thm:quantile_policy_gradient}
The objective gradient $\nabla_\theta J_{risk}(\alpha;\theta)$ has the same direction as  $E[D(\tau; \theta, q(\alpha;\theta))]$, where
\begin{equation} 
    \begin{aligned}
        D(\tau; \theta, r) = -\mathbf{1}\{R(\tau_i)\leq r \} \sum_{t=1}^{T}\nabla_\theta \log \pi(a_t| s_t; \theta).
    \end{aligned}
\end{equation}
\end{theorem}
With Theorem \ref{thm:quantile_policy_gradient}, we can update the network parameters by
\begin{equation} \label{eq:quantile_policy_gradient}
    \theta_{i+1} = \theta_{i} + \gamma D(\tau_i;\theta_i,q_i).
\end{equation}

\subsection{Training Pipeline}
The MCTS and the policy optimization are executed alternately to mine alphas. Trajectories sampled from the MCTS will be the training data for the risk-seeking policy network. The trained policy network will be used in the selection and rollout procedures in the MCTS. In other words, the MCTS serves as a sampler to interact with the environment; using the sample trajectories, the policy optimization works as a optimizer to train a risk-seeking policy for the MCTS sampler. The policy network is composed of a GRU feature extractor and a multilayer perceptron policy head. The pseudocode for the mining pipeline is shown in Algorithm \ref{alg:pipeline}.

\section{Experiment}
We design experiments to answer the following questions:
\begin{itemize}
\item \textbf{Q1:} How does our proposed method performance compare with other state-of-the-art methods?
\item \textbf{Q2:} How do the synergistic alphas perform in a realistic trading scenario?
\item \textbf{Q3:} How do different components in our methods contribute to the overall performance?
\end{itemize}

\subsection{Experiment Setting}
\subsubsection{Data.} We evaluate our alpha mining pipeline on China's A share market. 
Six features on stock price and volume are used in our experiments: 
open, high, low, close, volume, vwap(volume-weighted average price). 
We use the raw stock feature as input and mine alphas that have a good correlation with future returns. 
The returns are computed on the close price and we have two targets for the returns. 
One is the {\bfseries 5-day} returns, the other is {\bfseries 10-day} returns. 
The dataset contains 13 years of daily data and is split into 
a training set (2010/01/01 to 2019/12/31), a validation set (2020/01/01 to 2020/12/31), a test set (2021/01/01 to 2022/12/31). 
Two popular stock sets are used in the experiments: constituent stocks on CSI300 index and CSI500 index.

\subsubsection{Baselines.} To demonstrate the superiority of our method, we compare it with a variety of benchmarks. First, we compare with methods about formulaic alphas. 
\begin{flushleft}
\textit{Formulaic alphas:}
\end{flushleft}
\begin{itemize}
\item {\bfseries Alpha101} is a list of 101 formulaic alphas widely known by industrial practitioners. To ensure the fairness of the experiment, the 101 alphas are combined linearly to form a mega-alpha in the experiment.
\item {\bfseries Genetic Programming} is a popular method of generating alpha by manipulating the structure of the expression tree. Most previous alphas generating methods are based on GP which generates one alpha at a time using IC as fitness measures. We use implementation by  {\bfseries gplearn} in the experiment. Alphas generated by gplearn also are combined linearly to generate a mega-alpha.
\item {\bfseries Alphagen} is a new framework for generating synergistic alphas by reinforcement learning. It is the current SOTA of generating formulaic alpha. We use the authors' official implementation in our experiment.
\end{itemize}

To better evaluate the effectiveness of our method, we also compare it with direct forecasting machine learning models implemented on qlib\cite{qlib}. For each stock, the models look back 60 days with the 6 price/volume features to construct 360-dimension input data. The models are trained by regression on future returns and then give the predictive score. The hyperparameters of these end-to-end models are set according to the benchmarks on qlib.

\begin{flushleft}
\textit{Neural network model:}
\end{flushleft}

\begin{itemize}
\item {\bfseries MLP}: a fully connected neural network of interconnected nodes that process input data.
\item {\bfseries GRU}: Gated Recurrent Unit, a type of recurrent neural network architecture, is particularly effective for processing sequential data.
\end{itemize}

\begin{flushleft}
\textit{Ensemble model:}
\end{flushleft}
\begin{itemize}
\item {\bfseries XGBoost}: a highly efficient and popular machine learning algorithm that uses a gradient boosting framework\cite{xgb}.
\item {\bfseries LightGBM}: a fast, distributed, high-performance gradient boosting framework based on decision tree algorithms\cite{lgbm}.
\item {\bfseries CatBoost}: an open-source, gradient boosting toolkit optimized for categorical data and known for its robustness and efficiency\cite{catboost}.
\end{itemize}

\subsubsection{Evaluation Metrics.} We use industry-wide accepted metrics to evaluate the performance of the algorithm. 
\begin{itemize}
\item {\bfseries IC}: a statistical measure expressing the correlation between predicted score and actual stock returns.
\item {\bfseries ICIR}: a performance metric that measures the consistency of the predictive ability over time by dividing the IC by its standard deviation.
\item {\bfseries RankIC}: the correlation between the predicted ranks and the actual ranks of asset returns, offering a robust approach to evaluating forecasts that is less sensitive to outliers.
\end{itemize}

\input{Tables/main_result}
\subsection{Main Results}

To answer \textbf{Q1}, we conduct experiments on different methods including formulaic alphas generator(gplearn, alphagen) and direct stock trend forecasting model(MLP, GRU, XGBoost, LightGBM, CatBoost). A widely known alpha set, alpha101, is also included in the experiment as a benchmark.

Table \ref{tab:main_results} shows our signal-based main results. Our proposed method achieves the best results across all baselines. The synergistic alphas discovered by RiskMiner not only exhibit strong predictive ability(high IC and RankIC) but also have stable performance(high ICIR). The alphagen comes in the second tier. It has competitive results on IC and RankIC. However, the gap in the ICIR indicator shows that the predictive power of the alphas generated by alphagen is not stable enough compared with our method. The end-to-end forecasting models have moderate performance in the experiment since they use raw stock data to predict future trends without elaborate feature engineering. The alphas generated by traditional GP are no longer competitive with other methods generating formulaic alpha. The performance of the well-known alpha101 has decayed dramatically since its discovery.

\begin{figure}[h]
    \centering
    \includegraphics[width=\linewidth]{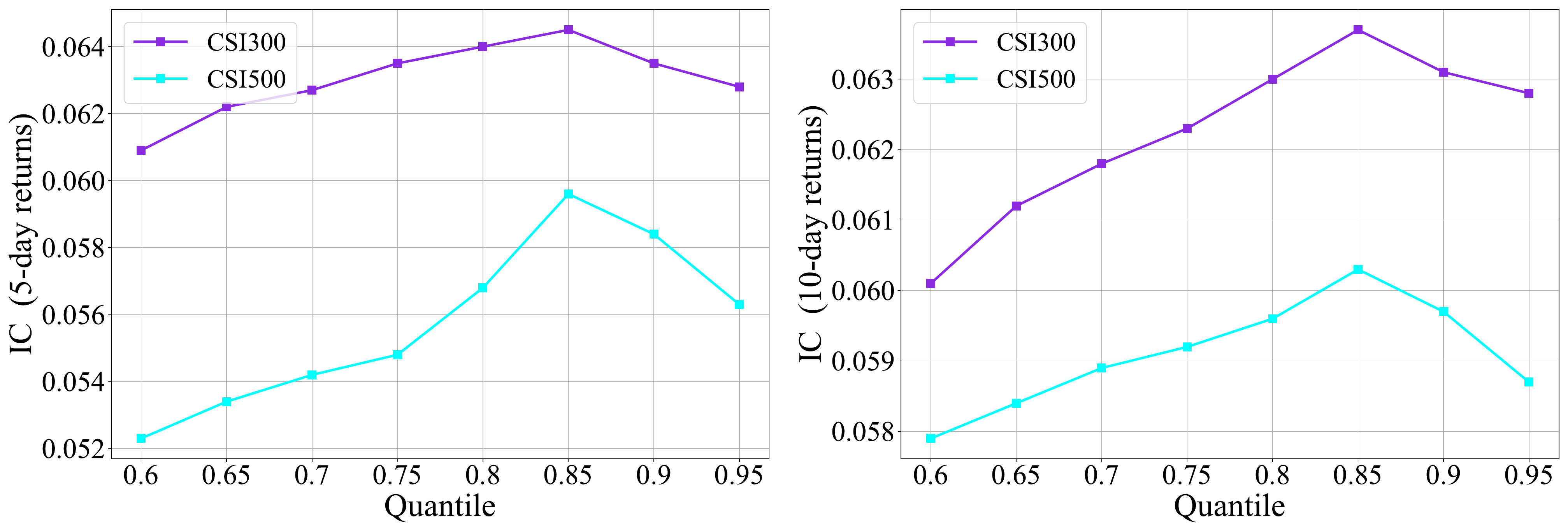}
    \caption{Model performance under different quantile level}
    \label{fig:IC_quantile_day}
\end{figure}

Since using quantile as the risk measure to facilitate the search process, it is necessary to investigate how the model performs under different risk-seeking levels. We run the mining pipeline under quantile values of $\alpha \in \{0.6, 0.65, 0.7, 0.75, 0.8, 0.85, 0.9, 0.95\}$ and monitor the IC performance. The results are shown in Figure \ref{fig:IC_quantile_day}.

In the beginning, the IC scores increase as the quantile level increases. It indicates that the risk-seeking policy can efficiently discover high-quality alphas since it directly optimizes the best-case performance. However, when the quantile value exceeds 0.85, the performance of our pipeline starts to decline. Therefore, an overly aggressive search strategy does not lead to significant improvements in results. On the contrary, it can lead to performance degradation. Selecting the appropriate quantile is crucial to the experimental outcomes.

A plausible explanation for this phenomenon could be that: even though the risk-seeking policy is more likely to discover better alphas, an overly aggressive policy may stuck in some local optimum. \emph{Mining alphas factor is different from other traditional RL tasks. We want to search as much as local optimums as possible instead of finding one global optimum.} As the risk-seeking level increases, the model would spend most of the search budget on the current best local optimum and fail to investigate other possible optimums in the expression space.

\subsection{Backtesting Results}
\begin{figure*}[!t]~
  \includegraphics[width=\textwidth]{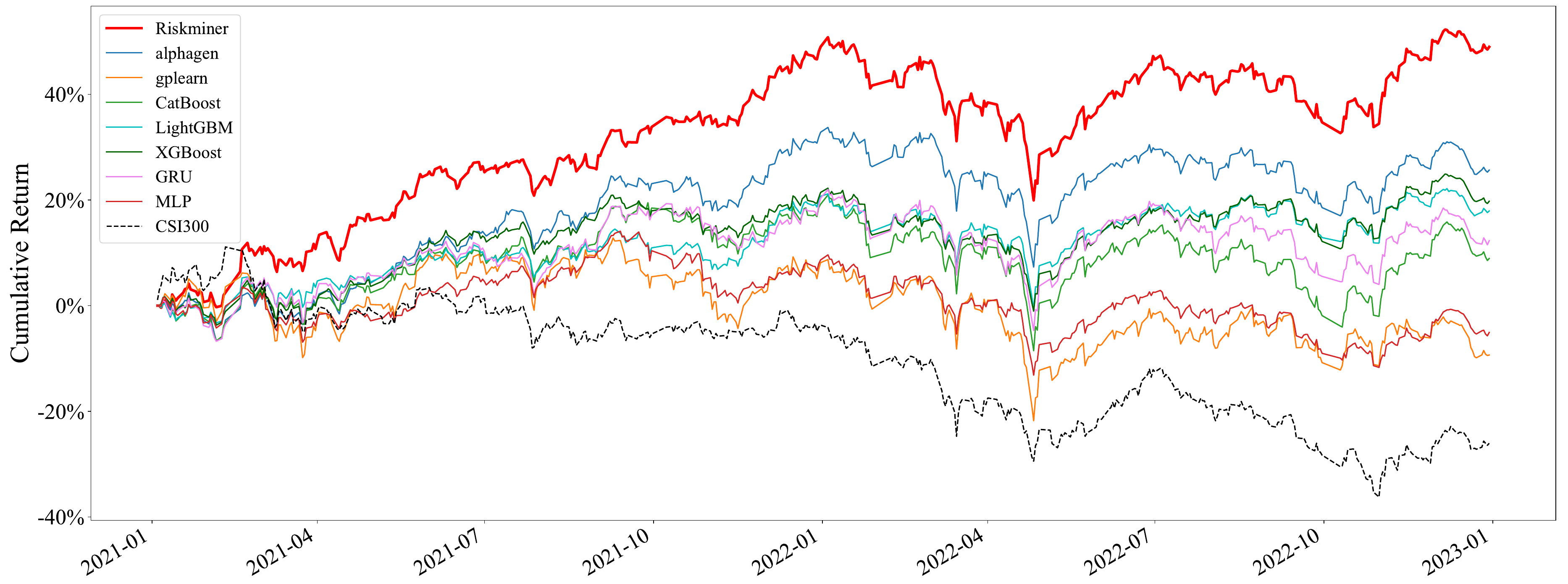}
  \caption{Backtest results on CSI300. The lines represent the cumulative return of agents under different alpha mining methods.}
  \label{fig:backtest_result}
\end{figure*}

To answer \textbf{Q2}, we conduct a simulated trading experiment on the test set (from 021/01/01 to 2022/12/31). We use the 5-day alphas in the previous experiment as trading signals and we rebalance our stock position weekly(every 5 days). We have excluded stocks that hit the price limit up or down and those suspended from trading. We adopt a simple long strategy in the experiment. Specifically, we rank the stock on day $t$ according to the predicted score or alpha value. Then we select the top $k$ stocks to evenly invest in them and sell currently held stocks that rank lower than $k$. We use cumulative returns as the portfolio metrics to evaluate the trading performance. 

\begin{itemize}
\item {\bfseries Cumulative return}: cumulative return is defined as the total change in the value of an investment or portfolio over a set period.
$$\mathrm{CR} = \frac{\mathrm{Final\:Value}}{\mathrm{Initial\:Value}} - 1$$
\end{itemize}

To find the best parameter $k$, we conduct grid searches on the validation set. The candidate parameter set for $k$ is $\{10, 20, 30, 40, 50, 60\}$. We discover that $k=40$ can maximize the cumulative returns on the validation set. The results are shown in Figure \ref{fig:backtest_result}. It is worth noting that the A-share market was in a bear market from 2021/01/01 to 2022/12/31. During the bear market, deriving profitable long strategies is extremely challenging. Certain models, including gplearn and MLP, incur losses in the period. Alphas by alphagen have desirable performance compared with other methods. However, our proposed RiskMiner outperforms the alphagen by a large margin.

\subsection{Ablation Study}

\input{Tables/ablation}
To answer \textbf{Q3}, we conduct the ablation study on the CSI300 dataset. Note that MCTS and the risk-seeking policy can work separately as alphas generators to search the expression space from the alphapool's reward. We investigate the individual performance of the two modules and the aggregate performance when they are working together. The results are shown in Table \ref{tab:ablation}. The performance of individual modules degrades to a certain extent. The MCTS has a slight advantage over the Risk-seeking policy. We can infer that MCTS contributes significantly to the overall performance. By integrating with a risk-seeking policy, our methods can elevate performance to a new tier.

\section{Conclusion}
In this paper, we propose RiskMiner, a novel framework for generating synergistic formulaic alphas. A reward-dense MDP is designed to stabilize the search. Then we integrate MCTS with the risk-seeking policy enabling Riskminer to effectively navigate through vast discrete solution space in alpha mining. We demonstrate the effectiveness of RiskMiner through extensive experiments on real-world datasets, showcasing its superiority in discovering synergistic alphas with strong predictive abilities and stable performance. The proposed method outperforms all existing state-of-the-art methods. Moreover, we give insights into the importance of selecting appropriate risk-seeking levels to avoid performance degradation, which provides valuable guidance for application in the industry. In the future, a possible direction for our research is constructing sentimental alphas with the strong text-processing ability of LLM.

\begin{acks}
This work was supported in part by the National Natural Science Foundation of China (NSFC) under Grants 72325007,72250065,72022001
\end{acks}

\bibliographystyle{ACM-Reference-Format}
\bibliography{main}

\appendix

\section{Proof  of Theorem \ref{thm:quantile_policy_gradient}}
\begin{proof}
By the definition of the $\alpha$-quantile, we have $$F_R(q(\alpha;\theta);\theta) = \alpha.$$
With the implicit function theorem, we can obtain
\begin{equation*}
    \nabla_{\theta}q(\alpha;\theta) = -\frac{\nabla_{\theta}F_R(r;\theta)}{f_R(r;\theta)}\bigg|_{r=q(\alpha;\theta)},
\end{equation*}
where $f_R(\cdot;\theta)$ is the non-negative density function and implies
\begin{equation*}
    \nabla_{\theta}q(\alpha;\theta) \ \propto -\nabla_{\theta}F_R(r;\theta)|_{r=q(\alpha;\theta)}.
\end{equation*}
Since $\mathbf{1}\{R(\tau)\leq r\}$ is an unbiased estimator for $F_R(r;\theta)$, we derive the CDF gradient as follows:
\begin{equation*}
    \begin{aligned}
        \nabla_{\theta}F_R(r;\theta)
        & = \nabla_{\theta} \mathbb{E}[\mathbf{1}\{R(\tau)\leq r\}] = \nabla_{\theta}\int_{\Omega_{\tau}}\mathbf{1}\{R(\tau)\leq r\}\Pi(\tau;\theta)d\tau\\
        & = \mathbb{E}[\mathbf{1}\{R(\tau)\leq r\} \nabla_{\theta} \log \Pi(\tau;\theta)],
    \end{aligned}
\end{equation*}
where $\Omega_{\tau}$ is the trajectory space, and the third equality comes from the likelihood ratio technique.
Note that $\nabla_\theta\log \Pi(\tau;\theta)=\sum_{t=1}^{T}\nabla_\theta\log\pi(a_{t}|s_{t-1};\theta)$. We further have
\begin{align*}
        \nabla_{\theta}F_R(r;\theta) & = \mathbb{E}[\mathbf{1}\{R(\tau)\leq r\} \sum_{t=1}^{T} \nabla_{\theta} \log \pi(a_{t}|s_{t-1};\theta)].
\end{align*}
Thus, we obtain the unbiased estimator for $-\nabla_{\theta}F_R(r;\theta)$ as follows:
\begin{equation*}
    D(\tau; \theta, r) = -\mathbf{1}\{R(\tau_i)\leq r \} \sum_{t=1}^{T}\nabla_\theta \log \pi(a_t| s_{t-1}; \theta),
\end{equation*}
which completes the proof.
\end{proof}

\section{Pseudo Codes}
we provide pseudocode for the alphapool and the overall alpha mining framework.
\begin{algorithm}
\caption{maintain the alpha pool}
\label{alg:alphapool}
    \KwIn{alpha set $\mathcal{F}=\{f_1,f_2,\dots,f_k\}$; a new alpha $f_{new}$; the alpha combination model $c(\cdot|\mathcal{F},\omega)$}
    \KwOut{optimal alpha set $\mathcal{F}^* = \{f^*_1,f^*_2,\dots,f^*_k\}$ and its weight $\omega^*=(\omega^*_1,\omega^*_2,\dots,\omega^*_k)$}
    $\mathcal{F}\gets \mathcal{F}\cup f_{new}$, , $w\gets w \Vert \mathrm{rand}()$;\\
    \For{each iteration}{
        Calculating $z_t=c(X_t|\mathcal{F},\mathbf{\omega})$ and $\mathcal{L}(\mathbf{\omega})$;\\
        $\omega \gets$ GradientDescent $(\mathcal{L}(\omega))$;\\
    }
    $i=\arg\min_i |\omega_i|$;\\
    $\mathcal{F}\gets\mathcal{F} /\{f_i\}$, $\mathbf{\omega}\gets\ (\omega_1,...,\omega_{i-1},\omega_{i+1},...,\omega_k)$;\\
    \Return $\mathcal{F}$ and $\omega$
\end{algorithm}
\begin{algorithm}
\caption{alpha mining pipeline}
\label{alg:pipeline}
    \KwIn{raw stock data $X_t$}
    \KwOut{optimal alpha set $\mathcal{F}^* = \{f^*_1,f^*_2,\dots,f^*_k\}$ and its weight $\omega^*=(\omega^*_1,\omega^*_2,\dots,\omega^*_k)$}
    Initialize $\mathcal{F}$ and $\omega$;\\
    Initialize $\pi_{risk}(\cdot|\theta)$ and replay buffer $\mathcal{B}=\{\}$;\\
    \For{each iteration}{
        Reset the root node of the search tree;\\
        Empty $\mathcal{B}$;\\
        \For{each interation}{
            Select by Equation \ref{eq:PUCT} to reach the leaf node then expand;\\
            Rollout  by $\pi_{risk}(\cdot|\theta)$ then backpropagation;\\
            Update $\mathcal{F}$ and $\omega$ using Algorithm \ref{alg:alphapool};\\
            $\mathcal{B} \gets \mathcal{B}\cup\tau$;\\
        }
        \For{each $\tau \in \mathcal{B}$}{
            Using Equation \ref{eq:quantile_update} to estimate current quantile;\\
            Using Equation \ref{eq:quantile_policy_gradient} to update policy network parameter;\\
        }
    }
    \Return $\mathcal{F}$ and $\omega$
\end{algorithm}

\section{Operators and Operands}
All the operators and operands used in our mining framework are listed. There are three types of operands: price/volume feature, times deltas, and constant. Times deltas can only be processed by the time-series operators. The operands are listed in Table \ref{tab:operands}. The operators can be divided into 2 groups: cross-section and time-series operators. In each category, operators can be further divided as unary and binary. The operators are listed in Table \ref{tab:operators}

\input{Tables/operands}

\input{Tables/operators}

\section{Implementation details}
We set the alphapool size $K=100$. The parameter $\lambda$ in the reward-dense MDP is set to 0.1. The GRU feature extractor has a 4-layer structure and the hidden layer dimension is 64. The policy head is MLP with two hidden layers of 32 neurons. In one mining iteration, the MCTS search cycle will be executed 200 times, and the sampled trajectories will be used for subsequent risk-seeking policy optimization. The learning rate $\beta$ for quantile regression is 0.01. The learning rate $\gamma$ for network parameter update is 0.001.

\end{document}

%% file: Tables/main_result.tex
\begin{table*}[!t]
  \caption{Main results on CSI300 and CSI500. All experiments repeat 10 times. Values outside the parenthesis are the mean and values inside the parenthesis are the standard deviation. All the evaluation metrics are the higher the better.}
  \label{tab:main_results}
  \begin{tabular}{c|cccccc|cccccc}
    \toprule
    \multirow{3}*{\textbf{Method}}& 
    \multicolumn{6}{c|}{\textbf{CSI300}} & \multicolumn{6}{c}{\textbf{CSI500}} \\
    \cline{2-13}
    &\multicolumn{3}{c}{5 days} & \multicolumn{3}{c|}{10 days}&\multicolumn{3}{c}{5 days} & \multicolumn{3}{c}{10 days} \\
    &IC&ICIR&RankIC&IC&ICIR&RankIC&IC&ICIR&RankIC&IC&ICIR&RankIC \\
    \midrule
    \multirow{2}*{MLP}&0.0273&0.1870&0.0396&0.0265&0.1772&0.0326&0.0259&0.1930&0.0389&0.0272&0.2042&0.0293 \\
    &(0.0042)&(0.0207)&(0.0052)&(0.0071)&(0.0258)&(0.0088)&(0.0021)&(0.0147)&(0.0039)&(0.0020)&(0.0160)&(0.0031) \\
    \multirow{2}*{GRU}&0.0383&0.2772&0.0584&0.0362&0.2883&0.0474&0.0378&0.2879&0.0576&0.0376&0.2987&0.0435 \\
    &(0.0031)&(0.0187)&(0.0031)&(0.0049)&(0.0150)&(0.0076)&(0.0017)&(0.0316)&(0.0027)&(0.0036)&(0.0117)&(0.0042) \\
    \hline
    \multirow{2}*{XgBoost}&0.0394&0.2909&0.0448&0.0372&0.3102&0.0423&0.0417&0.3327&0.0439&0.0368&0.3576&0.0479 \\
    &(0.0027)&(0.0219)&(0.0034)&(0.0038)&(0.0280)&(0.0054)&(0.0036)&(0.0408)&(0.0052)&(0.0050)&(0.0481)&(0.0067) \\
    \multirow{2}*{LightGBM}&0.0403&0.4737&0.0499&0.0398&0.3342&0.0512&0.0392&0.3481&0.0409&0.0416&0.3765&0.0537 \\
    &(0.0085)&(0.0190)&(0.0067)&(0.0052)&(0.0318)&(0.0060)&(0.0041)&(0.0270)&(0.0042)&(0.0032)&(0.0353)&
    (0.0044)\\
    \multirow{2}*{CatBoost}&0.0378&0.3714&0.0467&0.0431&0.4380&0.0583&0.0427&0.4529&0.0492&0.0386&0.4237&0.0426 \\
    &(0.0060)&(0.0306)&(0.0062)&(0.0067)&(0.0230)&(0.0078)&(0.0039)&(0.0523)&(0.0068)&(0.0056)&(0.0464)&(0.0057) \\
    \hline
    alpha101&0.0094&0.1107&0.0114&0.0127&0.1703&0.0165&0.0135&0.1875&0.0176&0.0116&0.1697&0.0189 \\
    \multirow{2}*{gplearn}&0.0283&0.2425&0.0298&0.0257&0.2653&0.0372&0.0327&0.3562&0.0463&0.0254&0.2938&0.0372 \\
    &(0.0089)&(0.0247)&(0.0073)&(0.0070)&(0.0190)&(0.0082)&(0.0059)&(0.0203)&(0.0112)&(0.0063)&(0.0291)&(0.0067) \\
    \multirow{2}*{alphagen}&0.0604&0.4023&0.0689&0.0593&0.3422&0.0633&0.0519&0.4296&0.0792&0.0537&0.5137&0.0708 \\
    &(0.0109)&(0.0288)&(0.0083)&(0.0090)&(0.0172)&(0.0120)&(0.0047)&(0.0260)&(0.0062)&(0.0052)&(0.0320)&(0.0065) \\
    \multirow{2}*{\textbf{RiskMiner}}& \textbf{0.0645} & \textbf{0.5126} &\textbf{0.0734}&\textbf{0.0637}&\textbf{0.5361}&\textbf{0.0728}&\textbf{0.0596}&\textbf{0.6420}&\textbf{0.0837}&\textbf{0.0603}&\textbf{0.5920}&\textbf{0.0752} \\
    &(0.0069)&(0.0270)&(0.0093)&(0.0083)&(0.0192)&(0.0107)&(0.0064)&(0.0183)&(0.0067)&(0.0043)&(0.0361)&(0.0055) \\
    
    \bottomrule
    
  \end{tabular}
\end{table*}

%% file: Tables/ablation.tex
\begin{table}
  \caption{Ablation study on CSI300.}
  \label{tab:ablation}
  \begin{tabular}{ccccc}
    \toprule
    \multirow{2}*{\textbf{MCTS}}&\multirow{2}*{\textbf{Risk-seeking policy}}&\multicolumn{3}{c}{5 days}\\
    &&IC&ICIR&RankIC\\
    \midrule
    $\checkmark$&$\checkmark$&\textbf{0.0645}&\textbf{0.5126}&\textbf{0.0734}\\
    $\times$&$\checkmark$&0.0617&0.4622&0.0693\\
    $\checkmark$&$\times$&0.0630&0.4501&0.0721\\
  \bottomrule
\end{tabular}
\end{table}

%% file: Tables/operands.tex
\begin{table}[h]
	\centering
	\caption{operands used in our framework.}
    \begin{tabularx}{\linewidth}{>{\hsize=.4\hsize}c|c}
        \toprule[1.5pt]
        \textbf{Operand} & \textbf{Description} \\
        \midrule
        Price/volume feature & open, high, close, low, volume, vwap \\
        Times deltas & 1, 5, 10, 20, 30, 40, 50 \\
        Constant & \makecell[c]{-30.0, -10.0, -5.0, -2.0, -1.0, -0.5, -0.01, \\0.5, 1.0, 2.0, 5.0, 10.0, 30.0} \\
        \bottomrule[1.5pt]
    \end{tabularx}
	\label{tab:operands}
\end{table}

%% file: Tables/operators.tex
\begin{table*}[h]
	\centering
	\caption{operators used in our framework.}
    \begin{tabularx}{\textwidth}{>{\hsize=.4\hsize}c|c}
        \toprule[1.5pt]
        \textbf{Operators} & \textbf{Description} \\
        \midrule
        $\mathrm{Sign}(x)$ & Return 1 if $x$ is positive, otherwise return 0.\\
        $\mathrm{Abs}(x)$& The absolute value of $x$. \\
        $\mathrm{Log}(x)$ & Natural logarithmic function on. $x$ \\
        $\mathrm{CSRank(x)}$ & \makecell[c]{The rank of the current stock’s feature value $x$ relative to the feature values of all stocks on \\ today’s date.}\\
        \hline
        $x+y$, $x-y$, $x \cdot y$, $x/y$  & Arithmetic operators. \\
        $\mathrm{Greater}(x, y)$, $\mathrm{Less}(x, y)$ & Comparison between two values.\\
        \hline
        $\mathrm{Ref}(x, t)$ & The value of the variable $x$ when assessed $t$ days prior to today. \\
        $\mathrm{Rank}(x, t)$ & \makecell[c]{The rank of the present feature value, $x$, compared to its values from today  going back up \\ to $t$ days.}\\
        $\mathrm{Skew}(x, t)$ & The skewness of the feature $x$ in past $t$ days prior to today. \\
        $\mathrm{Kurt}(x, t)$ & The kurtosis of the feature $x$ in past $t$ days prior to today. \\
        $\mathrm{Mean}(x, t)$, $\mathrm{Med}(x, t)$, $\mathrm{Sum}(x, t)$ & The mean, median, or total sum of the feature $x$ calculated over the past $t$ days. \\
        $\mathrm{Std}(x, t)$, $\mathrm{Var}(x, t)$ & The standard deviation or variance of the feature $x$ calculated for the past $t$ days. \\
        $\mathrm{Max}(x, t)$, $\mathrm{Min}(x, t)$ & The maximum/minimum value of the expression $x$ calculated on the past $t$ days. \\
        $\mathrm{WMA}(x, t)$, $\mathrm{EMA}(x, t)$ & \makecell[c]{The weighted moving average and exponential moving average for the variable $x$ calculated over \\the past $t$ days.} \\
        \hline
        $\mathrm{Cov}(x, y, t)$ & The covariance between two features $x$ and $y$ in the past $t$ days. \\
        $\mathrm{Corr}(x, y, t)$ & The Pearson's correlation coefficient between two features $x$ and $y$ in past $t$ days. \\
        \bottomrule[1.5pt]
    \end{tabularx}
	\label{tab:operators}
\end{table*}